\documentclass[pra,superscriptaddress,twocolumn,amsmath,amssymb]{revtex4-1} %floatfix
\usepackage{graphicx}
\usepackage{bm}
\usepackage{verbatim}
\usepackage{dcolumn}
\usepackage{color}
\usepackage{SIunits}
\usepackage{booktabs}
\usepackage{textcomp}
\usepackage{multirow}

\newcommand{\ket}[1]{\vert#1\rangle}
\newcommand{\bra}[1]{\langle#1\vert}
\newcommand{\e}[0]{\mathrm{e}}
\newcommand{\projector}[2]{\vert #1 \rangle\hspace{-0.3ex} \langle #2 \vert}

\newcommand{\bsigma}[0]{\boldsymbol\sigma}
\newcommand{\vect}[1]{{\mathbf #1}}
\newcommand{\uvect}[1]{\hat{{\mathbf #1}}}
\newcommand{\opdag}[1]{\hat{#1}^{\dagger}}

\newcommand{\braket}[2]{\langle#1\vert#2\rangle}

\newcommand{\mycite}[1]{~\cite{#1}}
\newcommand{\myciten}[1]{~\cite{#1}}

\begin{document}

\title{Quantum teleportation from a telecom-wavelength photon to a solid-state quantum memory}

\author{F\'elix Bussi\`eres} 
\email{felix.bussieres@unige.ch}
\author{Christoph Clausen}
\altaffiliation[Current address: ]{Present address: Vienna Center for Quantum Science and Technology, TU Wien - Atominstitut, Stadionallee 2, 1020 Vienna, Austria.}
\author{Alexey Tiranov}
\author{Boris Korzh}
\affiliation{Group of Applied Physics, University of Geneva, CH-1211 Gen\`eve 4, Switzerland}
\author{Varun~B.~Verma}
\author{Sae Woo Nam}
\affiliation{National Institute of Standards and Technology, Boulder, Colorado 80305, USA}
\author{Francesco Marsili}
\affiliation{Jet Propulsion Laboratory, California Institute of Technology, 4800 Oak Grove Drive, Pasadena, California 91109, USA}
\author{Alban Ferrier}
\author{Philippe Goldner}
\affiliation{Chimie ParisTech, Laboratoire de Chimie de la Mati\`ere Condens\'ee de Paris, CNRS-UMR 7574, UPMC Univ Paris 06, 75005 Paris, France}
\author{Harald Herrmann}
\author{Christine Silberhorn}
\author{Wolfgang Sohler}
\affiliation{Applied Physics / Integrated Optics Group, University of Paderborn, 33095 Paderborn, Germany}
\author{Mikael Afzelius}
\author{Nicolas Gisin}
\affiliation{Group of Applied Physics, University of Geneva, CH-1211 Gen\`eve 4, Switzerland}

\date{\today}

\begin{abstract}
In quantum teleportation\mycite{Bennett1993a}, the state of a single quantum system is disembodied into classical information and purely quantum correlations, to be later reconstructed onto a second system that has never directly interacted with the first one. This counterintuitive phenomenon is a cornerstone of quantum information science due to its essential role in several important tasks\mycite{Kok2007a,Briegel1998a,Kimble2008a} such as the long-distance transmission of quantum information using quantum repeaters\mycite{Sangouard2011a}. In this context, a challenge of paramount importance is the distribution of entanglement between remote nodes, and to use this entanglement as a resource for long-distance light-to-matter quantum teleportation. Here we demonstrate quantum teleportation of the polarization state of a telecom-wavelength photon onto the state of a solid-state quantum memory. Entanglement is established between a rare-earth-ion doped crystal storing a single photon that is polarization-entangled with a flying telecom-wavelength photon\mycite{Clausen2011a,Saglamyurek2011a}. The latter is jointly measured with another flying qubit carrying the polarization state to be teleported, which heralds the teleportation. The fidelity of the polarization state of the photon retrieved from the memory is shown to be greater than the maximum fidelity achievable without entanglement, even when the combined distances travelled by the two flying qubits is 25~km of standard optical fibre. This light-to-matter teleportation channel paves the way towards long-distance implementations of quantum networks with solid-state quantum memories.
\end{abstract}

\maketitle

Quantum teleportation\mycite{Bennett1993a} allows the transfer of a quantum state between remote physical systems through the use of quantum entanglement and classical communication. The combination of quantum teleportation with quantum memories provides scalable schemes for quantum computation\mycite{Kok2007a},
quantum repeaters\mycite{Briegel1998a,Sangouard2011a} and quantum networks\mycite{Kimble2008a}. 
Light-to-matter quantum teleportation was demonstrated by use of quantum memories based on warm\mycite{Sherson2006a} or cold\mycite{Chen2008a} atomic ensembles, or a quantum dot spin qubit\mycite{Gao2013a}. In these demonstrations, the memory emits a flying qubit (i.e.~a qubit encoded in a photon) with whom it is entangled, and the photon is used to distribute the entanglement necessary to perform teleportation.

In order to achieve long-distance light-to-matter quantum teleportation, and more generally to exchange quantum information between distant nodes of a quantum network in a scalable way,
we require efficient and multimode quantum memories with on-demand read-out, as well as 
a practical and efficient method to distribute entanglement\mycite{Sangouard2011a,Bussieres2013a}. Optical fibre is a naturally suited media for entanglement distribution, but it requires the flying qubits to have a suitable telecom wavelength (i.e.~with minimal absorption). 
Satisfying this requirement using the aforementioned emissive quantum memories is difficult, because the wavelength of the flying qubit is strictly determined by the discrete energy levels of the quantum memory, and this wavelength is typically far away from the low-loss region of standard optical fibre. 
Moreover, the spectral widths obtained with resonant emission in atomic ensembles are very narrow (a few MHz at most), which effectively reduces the rate at which the flying qubits can be generated, and thus the rate at which quantum teleportation can be attempted. Finally, these memories typically have a low multimode capacity\mycite{Afzelius2009a}, which limits the entanglement distribution rate in a long-distance configuration\mycite{Sangouard2011a}. 
To overcome these limitations, an approach based on sources of photon pairs combined with multimode quantum memories was proposed\mycite{Simon2007a}. The essential idea is that the sources create pairs comprised of one telecom-wavelength photon that is used to distribute entanglement between remote stations, while the other photon other is stored in a nearby quantum memory. In this context, quantum memories based on rare-earth doped crystals are promising candidates due to their large storage bandwidths and multimode capacity\mycite{Bussieres2013a}. In the recent years, they were used to demonstrate high-efficiency storage\mycite{Hedges2010a},
long coherence times\mycite{Heinze2013a}, 
multimode storage\mycite{Usmani2010a}, 
on-demand readout at the single-photon level\mycite{Timoney2013a}, storage of photonic entanglement\mycite{Clausen2011a,Saglamyurek2011a} and heralded entanglement between two crystals\mycite{Usmani2012a}.
Here we demonstrate quantum teleportation of the polarization state of a telecom-wavelength photon onto the state of a single collective excitation stored in a rare-earth-ion doped crystal. 
For this, a pair of broadband polarization-entangled photons is first generated from spontaneous parametric downconversion in nonlinear waveguides, and one photon from the pair is stored in a nearby rare-earth-ion doped crystal. The other telecom-wavelength photon from the entangled pair is sent to a Bell-state analyzer where it is jointly measured with a photon that is carrying the polarization qubit state to be teleported. The polarization state of the photon retrieved from the quantum memory is then analyzed with quantum state tomography, and the fidelity of several teleported states is shown to outperform the classical benchmark. We also performed teleportation in a configuration where the combined distance travelled by both telecom-wavelength photons is 25~km in standard optical fibre while still outperforming the classical benchmark, demonstrating the long-distance capability of our approach. Of crucial importance to achieve this is the use of highly-efficient superconducting nanowire single-photon detectors\mycite{Marsili2013a}, which significantly improves the success rate of the teleportation.

\begin{figure*}[!t]
\includegraphics[scale=0.8]{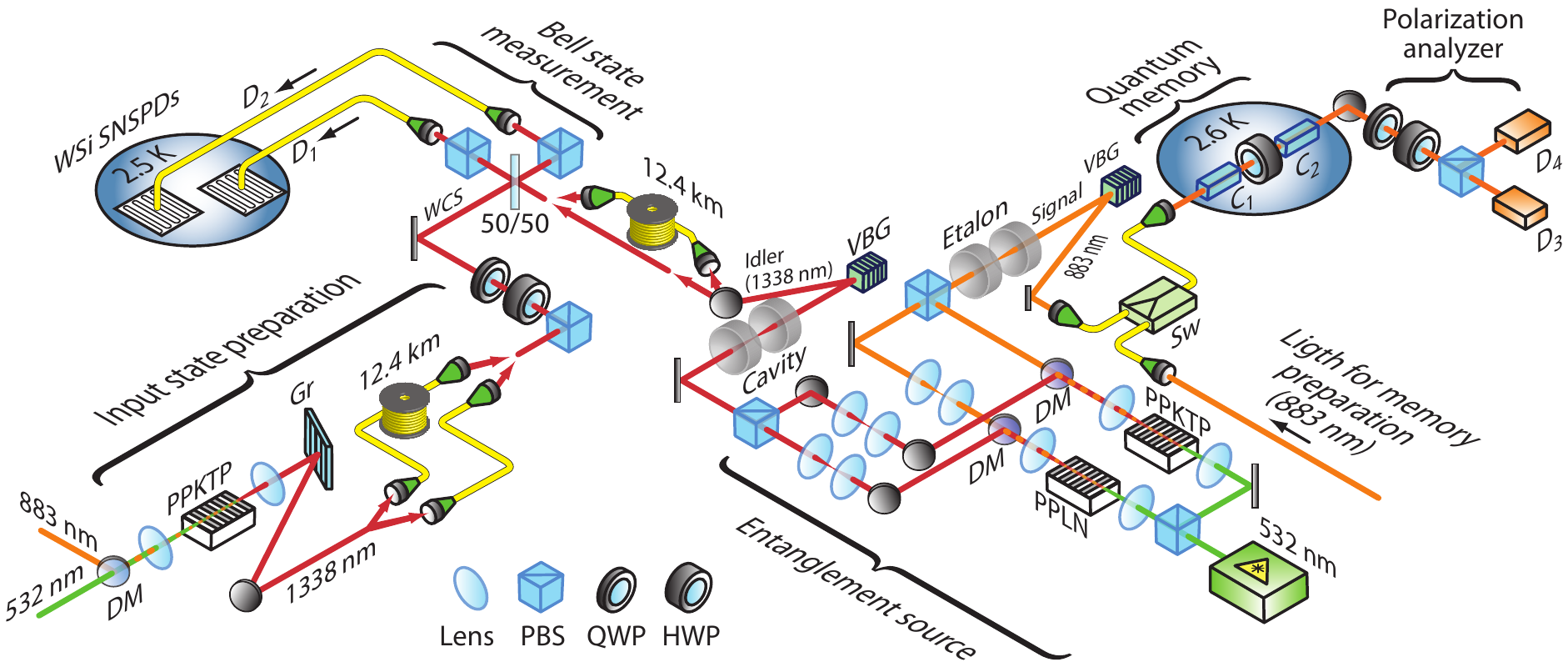}
\caption{\textbf{Experimental setup.} 
The system comprises the source of polarization-entangled photons at 883~nm (the signal) and 1338~nm (the idler) using filtered spontaneous parametric down conversion from two nonlinear waveguides (PPLN and PPKTP) coherently pumped with 532~nm light. After the waveguides, the signal and idler modes are separated using dichroic mirrors (DM) and are then individually manipulated to obtain good overlap after recombination at two polarizing beam splitters (PBS), as well as high transmission through the filtering cavity and etalon. A single pair of energy-correlated spectral modes of the signal and idler photons are selected using volume Bragg gratings (VBG). The signal photon is sent to a neodymium-based polarization-preserving quantum memory that was priorly prepared as an atomic frequency comb (AFC) using 883~nm light (see Methods). A switch (Sw) selects either the preparation light or the signal photons. The weak coherent state (WCS) at 1338~nm is created through difference-frequency generation from 532 and 883~nm light. The WCS is then selected using a grating (Gr) and coupled in an optical fibre. The input state to be teleported is prepared using wave plates and sent towards a 50/50 beam splitter (BS) where it is mixed with the idler photon to perform the Bell state measurement. The output modes of the BS are polarization filtered and sent towards two high-efficiency detectors based on WSi superconducting nanowires ($D_1$ and $D_2$) operated at 2.5~K in a closed-cycle cryocooler that is 10~m away from the quantum memory. A coincidence detection at $D_1$ and $D_2$ heralds a successful Bell state measurement. The signal photon retrieved from the quantum memory is sent to a polarization-state analyzer where it is detected on $D_3$ or $D_4$. The idler and WCS photons are each transmitted over either a short distance, or 12.4~km of single mode optical fibre.}
\end{figure*}

The experiment is represented on Fig.~1, and details are given in the methods and in the supplementary information (SI). A pair of entangled photons at 883~nm (the \emph{signal} photon) and 1338~nm (the \emph{idler} photon) is created from spontaneous parametric downconversion (SPDC). For this, 532-nm light is coherently pumping two nonlinear waveguides such that the photon pair is in superposition of being created in a first waveguide (with horizontal polarizations $\ket{HH}$) and in a second waveguide (with vertical polarizations $\ket{VV}$). Recombining the output modes of the waveguides on two polarizing beam splitters (PBS) yields two optical modes respectively containing the signal and idler photons prepared in an entangled state that is very close to $\frac{1}{\sqrt{2}}(\ket{HH}+\e^{i\varphi}\ket{VV})$. The spectra of the idler photon is afterwards filtered to a 240~MHz spectral width, corresponding to a coherence time $\tau = 1.4$~ns. Similarly, the signal photon is filtered to a width of 600~MHz with an etalon. However, due to energy conservation in SPDC, detection of the idler photon projects the signal's spectrum to a width that nearly reaches 240~MHz as well. 
This spectral width is more than five times larger than previous experiments with the same type of quantum memory\mycite{Clausen2011a,Usmani2012a}, which considerably increases the intrinsic repetition rate of our experiment.

Following the creation of a pair, the signal photon is directly sent to a 14~mm-long quantum memory consisting of two inline neodymium-doped yttrium orthosilicate crystals interspaced with a half-wave plate (HWP). This configuration compensates for the polarization-dependent absorption of a single crystal, and was previously shown to faithfully store arbitrary polarization states of single photons with a uniform efficiency\mycite{Clausen2012b,Gundogan2012a,Zhou2012a}. The absorption bandwidth of the quantum memory is of the order of 600 MHz and stores photons for 50~ns with an overall efficiency of 5\% using the atomic frequency comb (AFC) storage protocol\mycite{Afzelius2009a}. 
The qubit state to teleport, henceforth denoted the \emph{input state}, is encoded in the polarization of a photon from a weak coherent state (WCS) at 1338~nm that is created through difference-frequency generation in a separate nonlinear waveguide. This automatically yields the same central wavelength for the WCS and idler photons (see the SI). The indistinguishability of the WCS and the idler photon was checked independently of the quantum teleportation through the observation of a Hong-Ou-Mandel dip of 81\% visibility (see the SI).  
The Bell state measurement (BSM) between the idler photon and the input state is done by sending them through a 50/50 beam splitter, projecting their joint state on the Bell state $\ket{\Psi^-} = \frac{1}{\sqrt{2}}\left(\ket{HV}-\ket{VH}\right)$ when they are detected in different output modes\mycite{Zukowski1993a}. Two polarizers respectively selecting horizontal and vertical polarizations on those output modes remove accidental coincidences of photons with identical polarizations. The photons are then coupled in single mode optical fibres and detected using tungsten-silicide 
superconducting-nanowire single-photon detectors\mycite{Baek2011a,Verma2012a,Marsili2013a} (SNSPDs), shown as $D_1$ and $D_2$ on Fig.~1. These SNSPDs were specifically designed to operate at 2.5~K, which is higher than previous demonstrations (operating at around 1~K or less). This means the detectors could for the first time be operated in a simple two-stage closed-cycle cryocooler. Their efficiency depends on the bias current and reached 75\% with a temporal resolution (jitter) of $\sim 500$~ps. 
The jitter of the detectors is smaller than the coherence time of the photons, meaning that coincidences on the SNSPDs for which the WCS and idler photons overlap can be temporally resolved and post-selected. The teleportation is completed by retrieving the stored signal photon from the quantum memory and sending it into a polarization analyzer, where it is detected by single-photon detector $D_3$ or $D_4$. The qubit state of the retrieved photon, henceforth denoted the \emph{retrieved state}, requires a unitary correction\mycite{Bennett1993a} that is included in the polarization analyzer. 

In a first series of measurements, the WCS photon and the idler photon both travelled a few meters before the BSM (see Fig.~1), and their detection occurred while the signal photon was stored in the quantum memory. To post-select the threefold detections with the correct timing, we plot the temporal distribution of the measured threefold coincidences as a function of the delays $\delta t_{j1}$ and $\delta t_{j2}$ between a detection at $D_j$  ($j=3$ or $4$) and detections at $D_1$ and $D_2$. The results for the teleportation of the state $\ket{-} = \frac{1}{\sqrt{2}}(\ket{H}-\ket{V})$ are shown as two-dimensional histograms on Fig.~2-a (with $D_3$ projecting on $\ket{-}$), and 2-b (with $D_4$ projecting on $\ket{+}$).
Offsets on the detection times are chosen such that events at the vicinity of the centre of the histograms (i.e.~for $\delta t_{j1}, \delta t_{j2} <  \tau = 1.4$~ns) correspond to the actual teleportation (see SI).  
Fig.~2-a shows an increased number of counts at the centre, whereas Fig.~2-b has a dip, which is expected if the retrieved state is close to the input state $\ket{-}$. This is more easily visualized on Fig.~2-c (or 2-d), which shows a horizontal slice of 2-a (or 2-b) centred on $\delta t_{31} =0$ (or $\delta t_{41}=0$).

\begin{figure*}[!t] 
\includegraphics[scale=0.55]{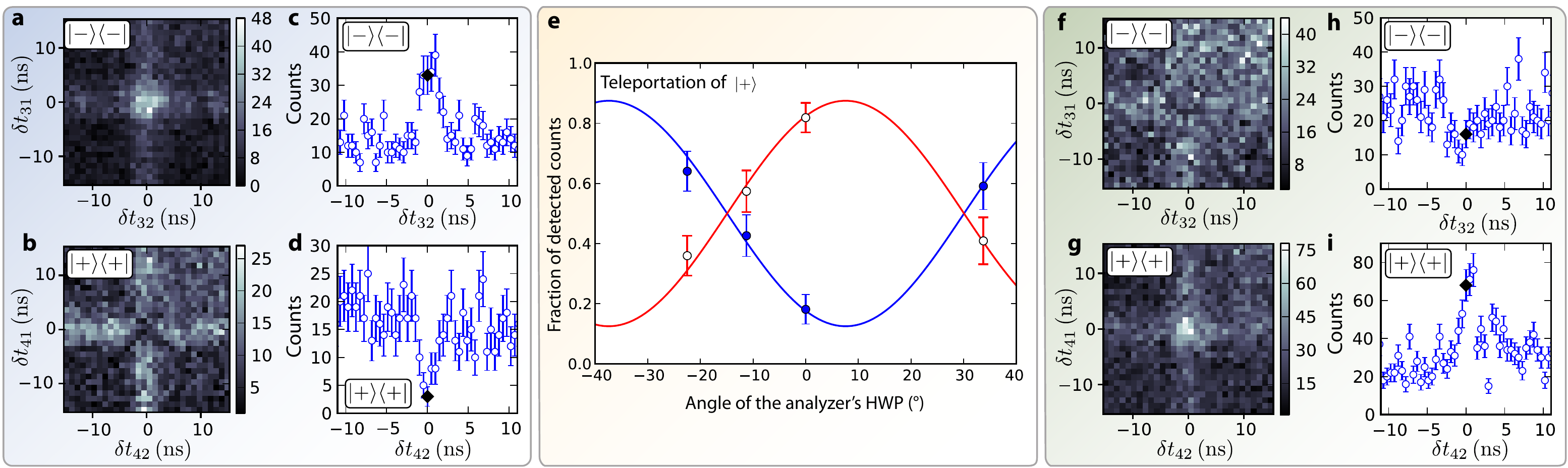}
\caption{\textbf{Experimental results.} 
The results of the teleportation of the input state $\ket{-}$ are shown on \textbf{a} through \textbf{d}. \textbf{a} is a two-dimensional histogram showing the number of threefold coincidences between detectors $D_1$, $D_2$ and $D_3$ as a function of the delays $\delta t_{31}$ and $\delta t_{32}$ between detections at $D_3$ and $D_1$ and $D_2$. \textbf{b} is the same as \textbf{a} with $D_4$ instead of $D_3$.  Each histogram indicates on which polarization state the retrieved photon was projected onto ($\projector{-}{-}$ for \textbf{a} and $\projector{+}{+}$ for \textbf{b}). Each pixel corresponds to a square time window whose side is of 486~ps duration. This is smaller than the coherence time of the photons, which is necessary to temporally resolve the detection events correspond a successful BSM.   
\textbf{c} and \textbf{d} are horizontal slices of \textbf{a} and \textbf{b} (centred on $\delta t_{31}=0$ and $\delta t_{41}=0$, respectively), and show the respective peak and dip in number of detections at the centre. The black diamonds shown are the points that have been used to estimate the fidelity of the teleportation. \textbf{e} shows the detected fraction of counts on $D_3$ and $D_4$ of the analyzer with input state $\ket{+}$, when the retrieved state is measured in a basis that is rotated around the equator of the Bloch sphere. Solid lines show the values expected from the quantum state tomography. \textbf{f} through \textbf{i} show the results of the teleportation of $\ket{+}$ when the combined distance travelled by the idler and WCS photons is 25~km of standard optical fibre.}
\end{figure*}

The fidelity of the retrieved state $\rho$ with respect to the input state $\ket{\psi}$ (which here is effectively pure), is $F = \bra{\psi}\rho\ket{\psi}$. It is equal to one if the teleportation is perfect, which implies that $\rho = \projector{\psi}{\psi}$. 
For the teleportation of the state $\ket{-}$ discussed above, the fidelity of the retrieved state can be readily be estimated from the number of events observed at the centre of Fig.~2-c ($\delta t_{32}=0$) and at the minimum of Fig.~2-d ($\delta t_{42}=0$), after a bias due to the different coupling and detection efficiencies of $D_3$ and $D_4$ is removed (see SI). 
The values that we used for the calculation of the fidelity are shown as the solid diamonds on 2-c and 2-d (see SI for an explanation of how these points have been selected). The measured fidelity is $92\pm 4\%$. To obtain complete information about the state $\rho$, we performed quantum state tomography\mycite{Altepeter2006a} by measuring in the $\{\ket{R},\ket{L}\}$ and $\{\ket{H},\ket{V}\}$ bases (see the SI for the histograms of these measurements). With this information, we can also calculate the purity $P = \text{Tr}(\rho^2)$ of the retrieved state, where $P=1$ for a pure state, $P=1/2$ for a completely mixed state, and $\frac{1}{2}<P<1$ otherwise. Since the input state is effectively pure, $P$ is related to the amount of depolarization caused by the teleportation process, which includes the storage of the signal photon. Here, the main cause of this depolarization is noise coming from multi-pair emission from the source and of multiple photons in the WCS; see the SI. The measured purity with the input state $\ket{-}$ is $94\pm 6 \%$. This value allows us to find an upper bound $F_{\text{max}} = \frac{1}{2}(1+\sqrt{2P-1}) = 97\pm 3\%$ on the observable fidelity, assuming the latter is affected by noise only, and not by an additional (and undesired) unitary rotation on the Bloch sphere.

\begin{table}
\begin{ruledtabular}
\caption{\textbf{Measured fidelities and purities} for all input states.  The uncertainties are obtained from Monte Carlo simulations assuming a Poisson distribution of the number of threefold events. Also shown is the upper bound on the fidelity  $F_{\text{max}}$ that is obtained from the measured purity.
}
  \begin{tabular}{@{} lccc @{}}
    %\hline
    Input state & Fidelity (\%) & Purity (\%) & $F_{\text{max}} (\%)$ \\ 
    \hline
    $\ket{H}$ & $94\pm 3$ & $93\pm 3$ & $96\pm3$\\ 
    $\ket{-} = \frac{1}{\sqrt{2}}(\ket{H}-\ket{V})$ & $92\pm 4$ & $94\pm 6$ & $97\pm 3$ \\
    $\ket{R}= \frac{1}{\sqrt{2}}(\ket{H}+i\ket{V})$ & $84\pm 4$ & $73 \pm 5$ & $84\pm 4$ \\
    $\ket{+} = \frac{1}{\sqrt{2}}(\ket{H}+\ket{V})$ & $82\pm 4$ & $83 \pm 9$ & $91 \pm 6$\\
    $\ket{+}$ (12.4~km) & $81\pm 4$ & --- & --- \\
    %\hline
  \end{tabular}
  \end{ruledtabular}
\end{table}

The fidelity and the purity of the retrieved state was also evaluated with other input states,  
and the results are listed in table~1 (see the SI for the histograms). The expected fidelity of an arbitrary state is $\bar{F}=\frac{2}{3}\bar{F}_e + \frac{1}{3}\bar{F}_p$, where $\bar{F}_e$ and $\bar{F}_p$ are the average fidelities measured on the equator and the pole, respectively. We find $\bar{F} = 89\pm4\%$, which is larger than the maximum fidelity of 66.7\% achievable with a prepare-and-measure strategy that does not use entanglement\mycite{Massar1995a}. The fidelities for states on the equator of the Bloch sphere ($\ket{+},\ket{-},\ket{R}$) are all smaller than for $\ket{H}$
which is consistent with the fact teleportation of the latter 
is unaffected by slow drifts of the phase $\varphi$ of the entanglement, and by the finite jitter of the detectors (see SI). However, the main factor limiting of the measured fidelities is the noise due to the creation of several pairs of photons and/or the presence of more than one photon in the WCS (see SI). Moreover, the observed differences in the purities indicate that this noise fluctuated from state to state.  
Finally, for the teleportation of $\ket{+}$, we show on Fig.~2-e the variation of the number of threefold coincidences when the measurement basis is gradually rotated around the equator of the Bloch sphere. The horizontal offset in the fitted curve is due to an additional rotation of the Bloch vector that, in addition to the purity-reducing noise, further reduces the observed fidelity below the $F_{\text{max}}$ upper bound. 

We also performed a teleportation of the $\ket{+}$ state in a configuration where the WCS photon and the idler photon each travelled through 12.4~km of standard single mode optical fibres before the BSM, yielding a combined travel distance of 24.8~km. The histograms show a dip (Fig.~2-f-h) and a peak (Fig.~2-g-i), which are indicatives of the teleportation. The fidelity of this measurement is $81\pm 4\%$, i.e.~the same as the one measured without the fibres. This is consistent with the fact that the added loss is roughly same for the WCS and idler modes, which keeps their relative contributions to the noise at the same level as without the fibres (see the SI). 

Our experiment demonstrates the feasibility of long-distance teleportation of a single quanta of light onto a solid-state quantum memory. The fundamentals of our experiment could be used to demonstrate a small-scale network of remote quantum memories, or a real-world quantum repeater based on an optical-fibre architecture. In this context, we require a quantum memory with on-demand readout based on storage on spin levels. Such levels are currently being used for spin-wave storage at the single-photon level in europium-doped crystals\mycite{Timoney2013a}, and are a promising avenue towards fulfilling requirements of a quantum repeater\mycite{Bussieres2013a}. 
Alternatively, the need for on-demand readout (in time) could be alleviated by exploiting 
spectral multimode storage in rare-earth crystals\mycite{Sinclair2013a}, which however needs to be complemented with a large increase in the number of other resources. In a broader context, our experiment could be useful to transfer quantum information between remote quantum network nodes made of rare-earth crystals coupled to superconducting qubits embedded in microwave resonators\mycite{Bushev2011a,Staudt2012a}, which could ultimately lead the realization of deterministic two-qubit gates\mycite{Steffen2013a} in solid-state network nodes. 

\begin{acknowledgements}
We thank Rob Thew, Pavel Sekatski and Hugo Zbinden for useful discussions. We acknowledge support by the European project QuReP and the Swiss NCCR QSIT. Part of the research was carried out at the Jet Propulsion Laboratory, California Institute of Technology, under a contract with the National Aeronautics and Space Administration. F.B.\@ and C.C.\@ contributed equally to this work.
\end{acknowledgements}

%\bibliography{teleportation}

\newpage
\section*{Methods}
\subsection*{WSi superconducting nanowire detectors}
The W$_x$Si$_{1-x}$ detectors optimized for maximum absorption at a wavelength of 1340 nm were fabricated on a 3 inch Silicon wafer. 
A gold mirror was fabricated by depositing 80~nm of gold on top of Ti using electron-beam evaporation and photolithographically patterned using a lift-off process. A space layer between the gold mirror and WSi detector consisting of 195~nm of SiO$_2$ was then deposited by plasma-enhanced chemical vapour deposition (PECVD). A~4.5 nm-thick W$_x$Si$_{1-x}$ layer ($x \approx 0.8$) was deposited by DC magnetron co-sputtering from separate W and Si targets at room temperature, and capped with 2 nm of amorphous Si to prevent oxidation. Electron-beam lithography and etching in an SF$_6$ plasma were used to define nanowire meanders with a width of 130~nm and pitch of 260~nm. An antireflection coating was deposited on the top surface consisting of 225~nm SiO2, 179~nm SiN$_x$, 231~nm SiO$_2$, and 179~nm SiN$_x$. A keyhole shape was etched through the Si wafer around each SNSPD, which could then be removed from the wafer and self-aligned to a single mode optical fibre\mycite{Marsili2013a}.  The size of the SNSPD is $16\times16$~\textmu m$^2$, larger than the 10~\textmu m mode field diameter of a standard single mode fibre, to allow for slight misalignment. When cooled to 2.5~K, the SNSPDs have an optimal system detection efficiency of 75\% with a dark count rate of the order of 300~counts per second. During the experiment, the temperature of the cryostat could fluctuate and the bias current would need to be adjusted accordingly to keep it sufficiently below the switching current. This could result in less than optimal performances such as, in the worst case, a detection efficiency of 60\% with a dark count rate of a few kHz.

\subsection*{Broadband source of polarization-entangled photons}
The source is using two periodically-poled (PP) nonlinear waveguides (see Fig.~1). One is a 1.3~cm-long waveguide embedded in potassium titanyl phosphate (PPKTP), and the other is a 6~cm-long titanium-indiffused waveguide based in lithium niobate (PPLN).  The fabrication of the PPLN waveguide required improvements of the poling technologies to realize periods as short as 6.5~\textmu m, which was required for this experiment. It comprises an input taper to facilitate the coupling of the pump light at 532~nm into the fundamental mode of the waveguide, and is coated with anti-reflection coatings. A pair of photons at 883 and 1338~nm is created from spontaneous parametric downconversion (SPDC). For this, pump light at 532~nm with 45\textdegree{} polarization is first split on a polarizing beam splitter (PBS). Waveguides are inserted in the two output modes of the PBS and are pumped coherently.  Hence a pair is in a superposition of being created in the PPLN waveguide (with both photons having horizontal polarization) and in the PPKTP waveguide (with both photons having vertical polarization). The pair creation probability is set to be low enough (at about 1\% within a time window of 500~ps) such that the creation of more than one pair is much less likely (with a probability $~10^{-4}$). 
The output modes of the waveguides a collimated, and the signal and idler modes are separated using dichroic mirrors. This yields four  spatial modes whose transverse profiles are individually manipulated with telescopes, whose necessity is described below. The two spatial modes of the signal photon are then recombined in a single one using a PBS. Similarly, the modes of the idler are also recombined using another PBS. After the two PBS, we have two spatial modes containing a pair of photons prepared in a state that is very close to $\frac{1}{\sqrt{2}}(\ket{HH}+\e^{i\varphi}\ket{VV})$ (see the SI). The telescopes are necessary to obtain a good overlap when recombining the two modes of the signal photon (likewise for the idler photon), and also to adjust the mode of the idler photon to match the mode of the Fabry-Perot cavity that comes after the PBS. The spectra of the idler (or signal) photon is filtered to a Lorentzian line width of 240~MHz (or 600 MHz) using the combination of a Fabry-Perot cavity (or an etalon) and a highly reflective volume Bragg grating (VBG). Several procedures are implemented to monitor and stabilize the properties of the source, which list here. Firstly, the pump light at 532~nm is continuously stabilized in frequency using a feedback mechanism based on difference-frequency generation of light at 1338~nm from mixing the 532 and 883~nm preparation light in the PPLN waveguide, as described in Ref.~\myciten{Clausen2011a}. This ensures that the energy of a pump photon is correlated with the central frequency of the spectra of the signal and idler photons, which are determined by the filters. Secondly, the residual 532~nm light present in the unused output ports of the two polarizing beam splitters that are just before the cavity and the etalon is used to continuously lock the phase $\varphi$ of the entangled state. For this, an error signal is derived from the 532~nm light, and a feedback is applied on two piezo mounted mirrors (one for the signal photon and one for the idler photon) that are placed right after the dichroic mirrors. Fast fluctuations are compensated for, but $\varphi$ could nevertheless slowly drift by a few degrees per hour, at most. Thirdly, a characterization of the properties of the source is performed every 30 minutes with a completely automatized procedure. For this, the teleportation is stopped momentarily for a few minutes by switching off the weak coherent state. Then, by measuring twofold coincidences between the idler photon and the transmitted signal photon, the visibility of the source is measured, and the value of the phase $\varphi$ is extracted (see SI). The measured average visibility was 93\%. From these measurements, the second-order cross-correlation function between the idler and the signal photons is estimated and used to monitor the probability $p$ to emit a pair of photons in a time window of about 500~ps. We measured $p\sim 10^{-2}$. The monitoring and stabilization yield stability for periods as long as 24~hours. We note that the 532-nm light is pulsed in 25~ns-long gaussian pulses with 100~ns between successive pulses, which is twice the storage time. This removes most of the detrimental optical noise stemming from the superposition of a signal photon retrieved from the memory with another photon created 50~ns later and that is transmitted through the memory instead of being stored\mycite{Usmani2012a}. This improves the signal-to-noise ratio of the teleportation by a factor $\ge 10$ with respect to previous experiments\mycite{Clausen2011a,Usmani2012a}. 

\subsection*{Polarization-preserving quantum memory}
The compact, broadband and polarization-preserving quantum memory was obtained by placing two 5.8~mm-long Nd$^{3+}$:Y$_2$SiO$_5$ crystals around a 2~mm-thick half-wave plate, resulting in a total device length of 14~mm. The light beam was focused in the centre of the crystals using a 250~mm lens. The compact device made it possible to achieve a sufficiently tight focus over the entire length, which is crucial for efficient optical pumping. Our previous memory device\mycite{Clausen2012b} required two separate passes through the cryostat, one for each crystal, thereby causing excessive losses due to the many optical surfaces. Here we also added anti-reflective coatings on all surfaces (cryostat windows, crystals and half-wave plate), further reducing losses. The resulting off-resonance transmission coefficient was 95\%, much higher than our previous devices. Using shorter crystals would, however, reduce the optical depth of the memory, which in turn would lower the intrinsic memory efficiency\mycite{Afzelius2009a}. To overcome this loss we grew customized Nd$^{3+}$:Y$_2$SiO$_5$ crystals using the Czochralski process, with higher neodymium concentration (estimated to be 75~ppm). These crystals have an optical inhomogeneous broadening of about 6~GHz, similar to our previous crystal (with 30~ppm doping level), but with a higher absorption coefficient of $\alpha=3.7$~cm$^{-1}$ (with an applied magnetic field of 300~mT). The resulting optical depth of the polarization-preserving memory device is $d= 2.3\pm0.1$. The optical pumping necessary for preparing the atomic frequency comb requires two ground-state levels, where one is used for population storage, which can be obtained by splitting the ground-state ${}^4I_{9/2}$ Zeeman doublet with a magnetic field. We used the same field strength and orientation as in previous memory demonstrations\mycite{Usmani2010a}. We measured a Zeeman population relaxation lifetime $T_Z=43$~ms in our new crystals, smaller than in the lower doped crystals (about 100~ms, see Ref.~\myciten{Usmani2010a}). The atomic frequency comb is prepared using an acousto-optic modulator (AOM), used in double-pass configuration, that modulates the intensity and frequency of the light from a external cavity diode laser at 883~nm (centred on the absorption line of the ${}^4I_{9/2}$-${}^4F_{3/2}$ transition) in order to pump some of the atoms to the other Zeeman level. This is used to create a 120~MHz comb with a spacing of 20~MHz between the peaks. To increase the memory bandwidth beyond the 120~MHz limit imposed by the effective bandwidth of the double-pass AOM, the light at the output of the AOM is sent into a phase modulator that creates first and second-order sidebands, all separated by 120~MHz. In this way, the comb at the carrier frequency is copied twice on each side, yielding an overall comb width of 600~MHz. The overall efficiency of the polarization preserving memory is 5\% with a 50~ns storage time. This is slightly lower than what we achieved in our previous device (about 8\%, see~\myciten{Clausen2012b}). We attribute this reduction to the shorter Zeeman relaxation time, which reduces the optical depth and contrast of the atomic frequency comb. The relaxation of the comb structure, during the 10~ms of the experimental cycle during which teleportation is performed, also causes a decay of the efficiency. Finally, the bandwidth extension using the sideband caused a comb structure with reduced contrast at the edges, due to insufficient optical power in the outer sidebands.

\newpage
\renewcommand{\thefigure}{S\arabic{figure}}
\renewcommand{\theequation}{S\arabic{equation}}
\renewcommand{\thetable}{S\arabic{table}}
%\renewcommand{\thesection}{A\arabic{section}}
%\renewcommand{\thesubsection}{A\arabic{section}-\arabic{subsection}}

%\appendix

\section*{Supplementary information}

\section{Introduction}
In this Supplementary Information, we provide additional details on our experiment. Section~\ref{section:monitoring} describes how the monitoring of the properties of the source of entangled photons and of the source of weak coherent state is done. Section~\ref{section:results} presents a detailed description of the features of the two-dimensional histograms of the threefold coincidences from which the quantum state tomography results are derived. We also model the noise stemming from multiple photon pairs and multiple photons in the weak coherent. Section~\ref{section:tomography} provides details on the how the quantum state tomography is performed. The effect of the aforementioned noise on the fidelity and purity is discussed.  

\section{Monitoring of the source of entangled photons and source of weak coherent state} \label{section:monitoring}
\subsection{Characterization of the source of entangled photon pairs} \label{section:sourcecharacterization}
The source of entangled photons was continuously monitored during the experiment. Here we provide details on how we monitored the relative phase $\varphi$ of the entangled state $\frac{1}{\sqrt{2}}(\ket{HH}+\e^{i\varphi}\ket{VV})$ that was produced, as well as the fluctuations of the number of photon pairs created in a given time window. 
\subsubsection{Entanglement visibility and phase drift compensation} 
Automatized monitoring of the source was performed a least once per hour by producing a visibility curve, which was accomplished as follows. First, the weak coherent state (WCS) was switched off (see Fig.~1), which is the same as Fig.~1 of the main text). Then, a half wave plate was inserted before the 50/50 beam splitter (BS) used for Bell state measurement, and its angle was set such that, when combined with the polarizers placed just after the beam splitter, a detection on $D_1$ would project on $\ket{+}$, and $D_2$ would project on $\ket{-}$. The state of the corresponding signal photon, e.g.~$\frac{1}{\sqrt{2}}\left|H\right\rangle+e^{i\varphi}\left|V\right\rangle$ when the detection occurred at $D_1$, was then analyzed by rotating the half wave plate (HWP) of the analyzer, projecting on states on the equator of the Bloch sphere. For this measurement we use only coincidences stemming from the \emph{transmitted photons}, i.e.~the signal photons that passed through the quantum memory without being absorbed.

\begin{figure}[!ht]
\begin{center}
\includegraphics[width=8.8cm]{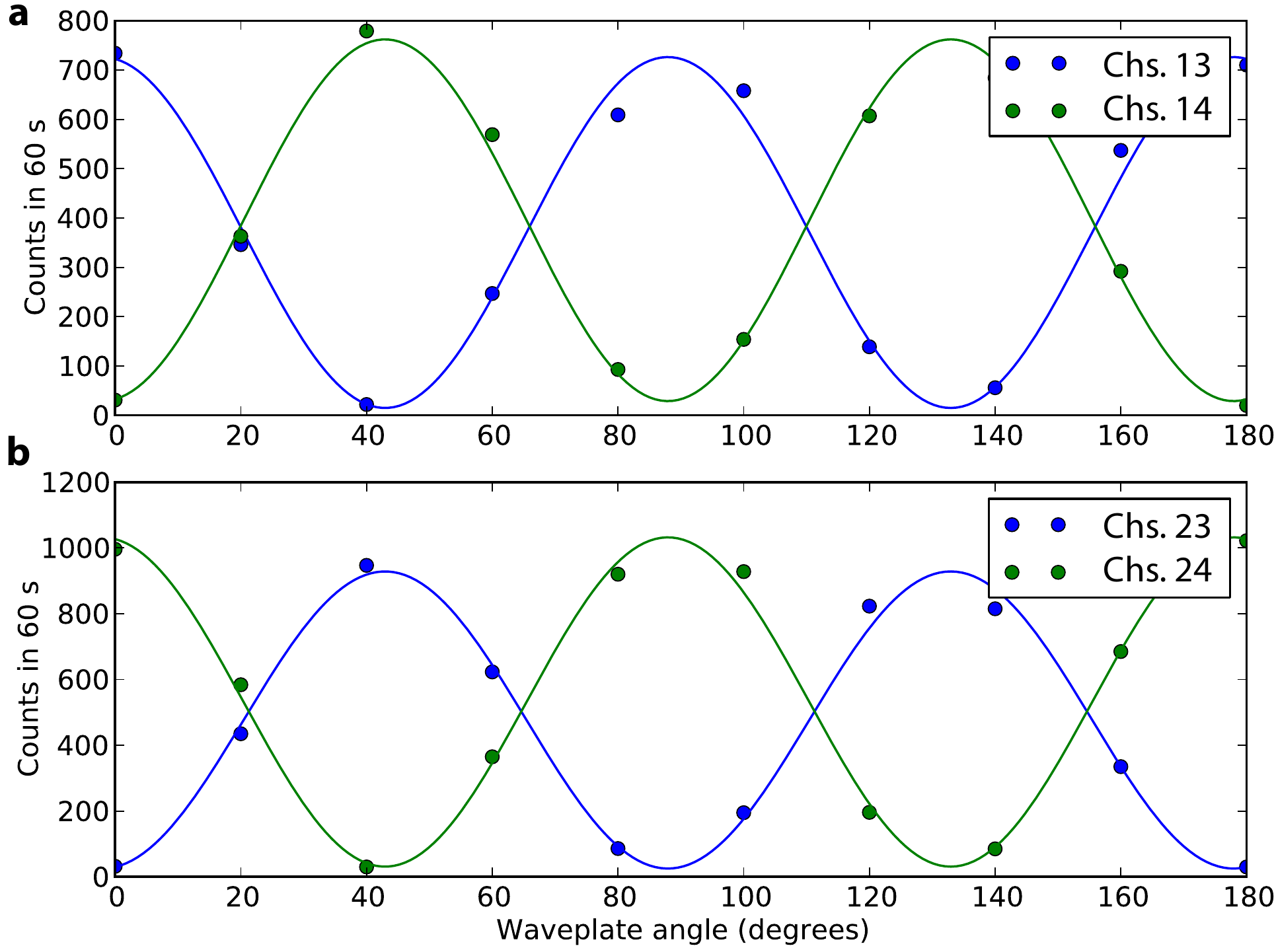}
\caption{Visibility curves for, \textbf{a}, detector pairs $D_1$-$D_3$ and $D_1$-$D_4$, and \textbf{b}, pairs $D_2$-$D_3$ and $D_2$-$D_4$, where $D_1$ and $D_2$ are the detectors for the idler photons, and $D_3$ and $D_4$ are the detectors of the analyzer of the signal photon.} \label{suppfig:visibility}
\end{center}
\end{figure}

The resulting visibility curves (see Fig.~\ref{suppfig:visibility} for an example) show the number of coincidences on each detector of the analyzer (i.e.~$D_3$ and $D_4$), as a function of the angle of the HWP. The phase $\varphi$ is determined from the common horizontal offset of the four curves. The phase slowly drifted with time, typically by a few degrees per several hours. For the teleportation measurements, this phase was effectively cancelled by rotating the quarter wave plate of the analyzer to set the offset of the visibility curves to zero. By monitoring the overall variations of the amplitudes of the visibility curves, we could also monitor the balance between two waveguides, as well as the coincidence rate of the source. The visibility, averaged over all measurements, was 93\%.

\subsubsection{Cross-correlation of idler and signal modes} \label{section:gsi}
To monitor the magnitude and stability of the number of photon pairs created in a given time window, we measured the zero-time second-order cross-correlation function between the detected idler photon and the transmitted signal photon, $g_{si}$, defined as
\begin{equation*}
 g_{si} = \frac{\langle d_i^\dagger d_i d_s^{\dagger} d_s
    \rangle}
  {\langle d_i^\dagger d_i \rangle\langle d_s^\dagger d_s \rangle},
\end{equation*}
where $d_i$ (or $d_s$) is the annihilation operator for the idler mode (or signal mode), and $d_i^{\dagger}$ (or $d_s^{\dagger}$) is the associated creation operator~\cite{Sekatski2012a}. With negligible dark counts and single-photon detectors having a timing resolution that is much smaller than the coherence time of the photons, one can show that $g_{si}= 1 + 1/p \approx p^{-1}$, where $p\ll 1$ is the probability to create a pair of photons in a given time window~\cite{Kuzmich2003a}. It is also equal to the ratio of the probability to detect a coincidence stemming from two photons of the same pair, over the probability to detect two photons from different pairs. Measuring a value $g_{si}> 2$ implies that the signal and idler fields are non-classically correlated~\cite{Kuzmich2003a}. Moreover, measuring a value $g_{si}\gg 1$ (which implies that $p\ll 1$) is a necessary condition to create close-to-maximally entangled states~\cite{Sekatski2012a} and to show the non-classical nature of the heralded signal photon~\cite{Clausen2012b}.

\begin{figure}[!t]
\begin{center}
\includegraphics[width=8.8cm]{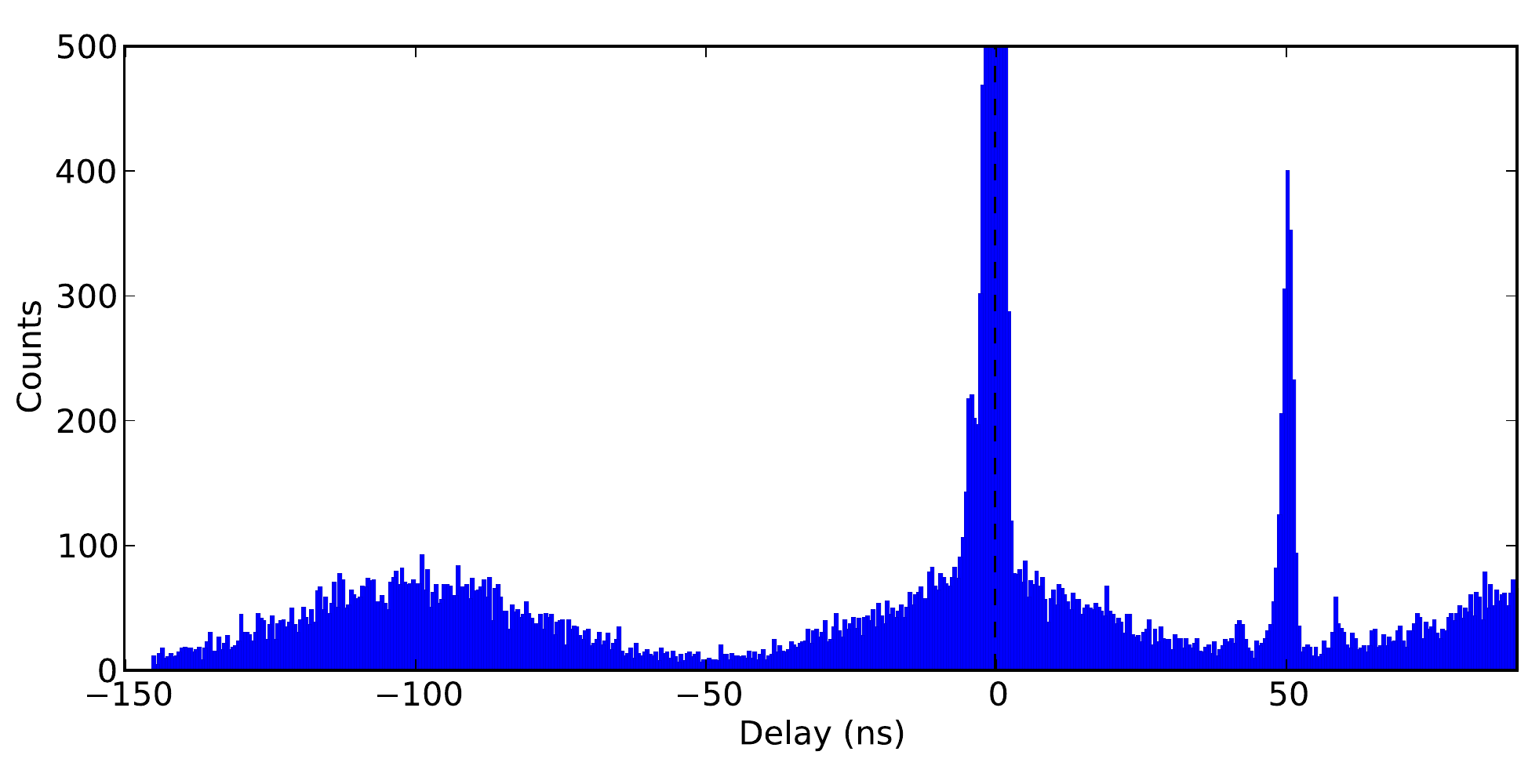}
\caption{Histogram of the number of coincidences as a function of the delay between the detection of a signal and an idler photon. Each bin is 162~ps wide. We see the peak corresponding to the detection of a transmitted signal photon (at 0~ns) and the peak of the stored and retrieved photon (50~ns). The transmitted peak is vertically clipped.} \label{suppfig:gsi}
\end{center}
\end{figure}

In practice, $g_{si}$ was estimated using all the data accumulated for the visibility curves during one day. Using these data, we produced a histogram of the number of coincidences between the idler and signal modes as a function of the delay between them. 
Fig.~\ref{suppfig:gsi}-a shows one such histogram, on which we can see two main peaks over an oscillating background level. 
The peak at 0~ns is due to coincidences involving a transmitted signal photon, and the one at 50~ns is due to a stored signal photon, i.e.~a signal photon that was stored and retrieved from the quantum memory. Note that the transmitted peak is clipped because we expanded the vertical scale so that we could see the effect of the pulsed pump laser, which gives rise to the wide and small bumps centred on -100, 0 and 100~ns, etc (we recall that the pump light at 25~ns was shaped into 25~ns-wide gaussian pulses separated by 100~ns). We see the reduction of the number of accidental coincidences between the bumps, which was the desired effect. The cross-correlation of the transmitted peak $g_{si}^{(t)}$ is estimated by dividing the number of coincidences in a narrow window centred on 0~ns by the number of coincidences in another window centred on a neighbouring bump that is 100~ns away. The average value of $g_{si}^{(t)}$ was 100, and varied from 80 to 150 for all the measurements. The cross-correlation of the stored photon $g_{si}^{(s)}$ is estimated by centering the first window on the stored photon peak, and the second one 100~ns away, which is falls on a minimum of the oscillating background. The measured value of $g_{si}^{(s)}$ varied from 6 to 20. All values were measured with 486~ps-wide coincidence windows. 
The measured values of $g_{si}^{(s)}$ fluctuate strongly, but they are nevertheless well above the classical upper bound of~2, which highlights the single-photon nature of the polarization state that is retrieved from the quantum memory~\cite{Kuzmich2003a,Usmani2012a,Clausen2012b}.

\subsection{Weak coherent state (WCS)}
As explained in the main text, the source of entangled photons was designed such that central frequency of signal photons corresponds to the centre of the atomic frequency comb that is created using the 883~nm diode laser, and such that the frequency of the pump light at 532~nm create photon pairs that satisfy the energy-conservation imposed by the transmission wavelength of the Fabry-Perot cavity of the idler photon. Hence, mixing part of the 532~nm light and part of the 883~nm diode laser into a separate PPKTP waveguide automatically creates coherent pulses of light (through difference-frequency generation, DFG) having a frequency that matches the central frequency of the Fabry-Perot cavity, and thus of the idler photons (see Fig~1). This light therefore has suitable spectral properties to be indistinguishable from the idler photons, and therefore to encode the input state of the teleportation. The intensity of the WCS was monitored and stabilized by diverting a small portion towards a single-photon detector creating a feedback signal controlling a variable attenuator. We estimated that the mean number of photons contained in a 486~ps-wide window at the centre of one WCS was $\mu \approx 0.011 \pm 0.002$ for the teleportation of $\ket{-}$, and $0.016$ for the teleportation of $\ket{+}$, $\ket{R}$ and $\ket{H}$.

\subsection{Indistinguishability of the idler and the WCS}
\begin{figure}[!t]
\begin{center}
\includegraphics[width=8cm]{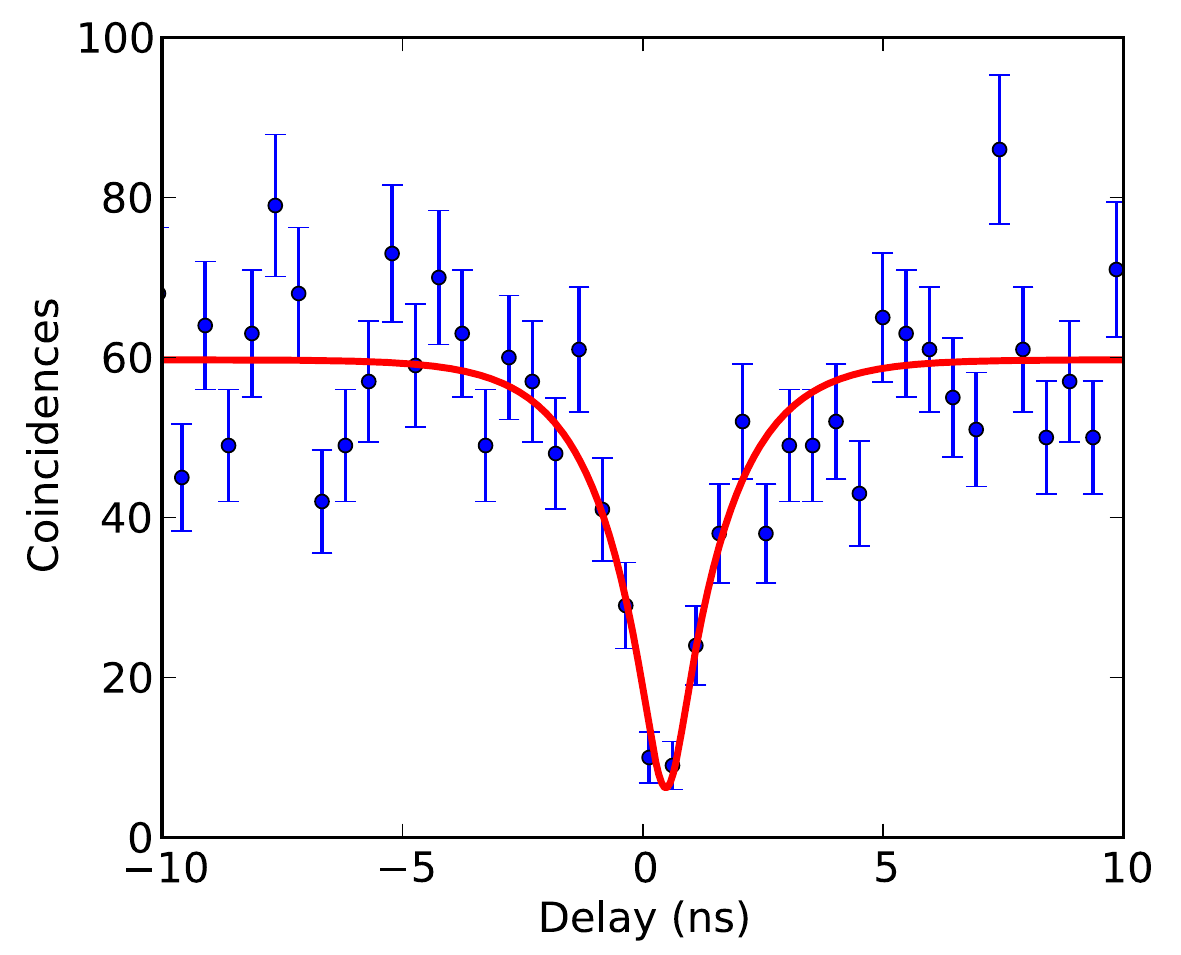}
\caption{Coincidence histogram showing a Hong-Ou-Mandel dip between an heralded signal photon and the weak coherent light. The visibility of the dip is 81\%. The horizontal axis is the delay between the detection of the signal photon and one of the detectors behind the 50/50 beam splitter.} \label{suppfig:hom}
\end{center}
\end{figure}

Projecting the input state and the idler photon on a Bell state (see Fig.~1) requires the ability to post-select events where the two photons temporally overlapped on the 50/50 beam splitter (see Fig.~1). This is possible only if the temporal resolution (i.e.~the jitter) of the detectors is smaller than the coherence time of the idler photon (because the WCS is generated from DFG between two narrowband lasers, its coherence time is much longer than the 1.4~ns coherence time of the idler). The temporal resolution effectively defines temporal modes on which the photons are projected onto when they are detected. Therefore, we need to consider the indistinguishability in these modes, which was verified through the observation of a Hong-Ou-Mandel dip in an experiment performed before the quantum teleportation~\cite{Hong1987a}. For this, continuous-wave (CW) light at 532~nm was used to pump the PPLN waveguide of the source while the PPKTP waveguide was blocked (see Fig.~1), and the filtered idler photons were mixed on the 50/50 beam splitter with the WCS. The signal photon was bypassing the quantum memory and used to herald an idler photon with an horizontal polarization, the same as the WCS. The idler photon was detected with a niobium nitride SNSPDs (7\% efficiency) that had jitter of about 100~ps. The photon-pair creation probability $p \approx 1/g_{si}$ was $\approx 0.0025$ in a 486-ps window, and the mean number of photon for the WCS was $\mu \approx 0.0035$. Fig.~\ref{suppfig:hom} shows the observed dip, with a visibility of 81\%. 
From this, we conclude that the idler photons and the WCS are close to be completely indistinguishable. The visibility is partly reduced by the noise stemming from the accidental detection of two photon from the WCS or two idler photons coming from two pairs created simultaneously. 

\section{Threefold detection histograms} \label{section:results}

\subsection{Teleportation of $\ket{-}$, $\ket{R}$ and $\ket{+}$}
\begin{figure}[!t]
\begin{center}
\includegraphics[width=7cm]{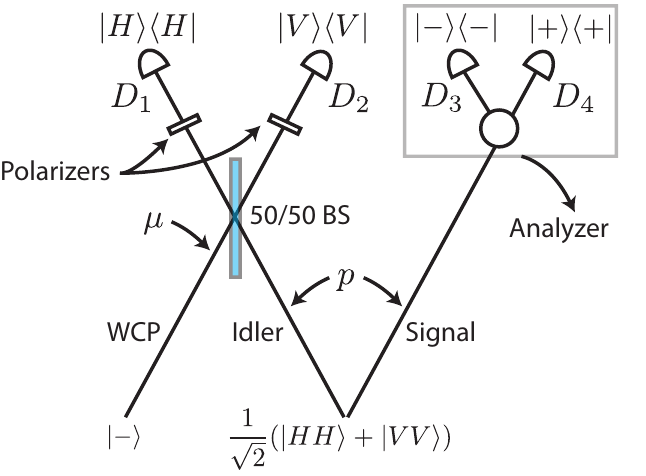}
\caption{Conceptual experimental setup for the teleportation of the $\ket{-}$ state. The polarization of a photon in the WCS is prepared in the the $\ket{-}$ state, and the probability to find a photon in a given time window is $\mu$. The idler and signal modes are populated with a pair of polarization-entangled photons with a probability $p$. The WCS and the idler modes are mixed on a 50/50 beam splitter (BS). The output modes of the BS are filtered with polarizers, such that a detection at $D_1$ projects on $\ket{H}$, and a detection at $D_2$ projects on $\ket{V}$. The signal mode is sent towards a polarization qubit analyzer set such that a successful Bell state measurement should result in a detection of the signal photon in detector $D_3$ (we represent this by indicating that $D_3$ projects on $\ket{-}$, and $D_4$ on $\ket{+}$). } \label{suppfig:simple-model}
\end{center}
\end{figure}

\begin{figure*}
\begin{center}
\includegraphics[width=16.5cm]{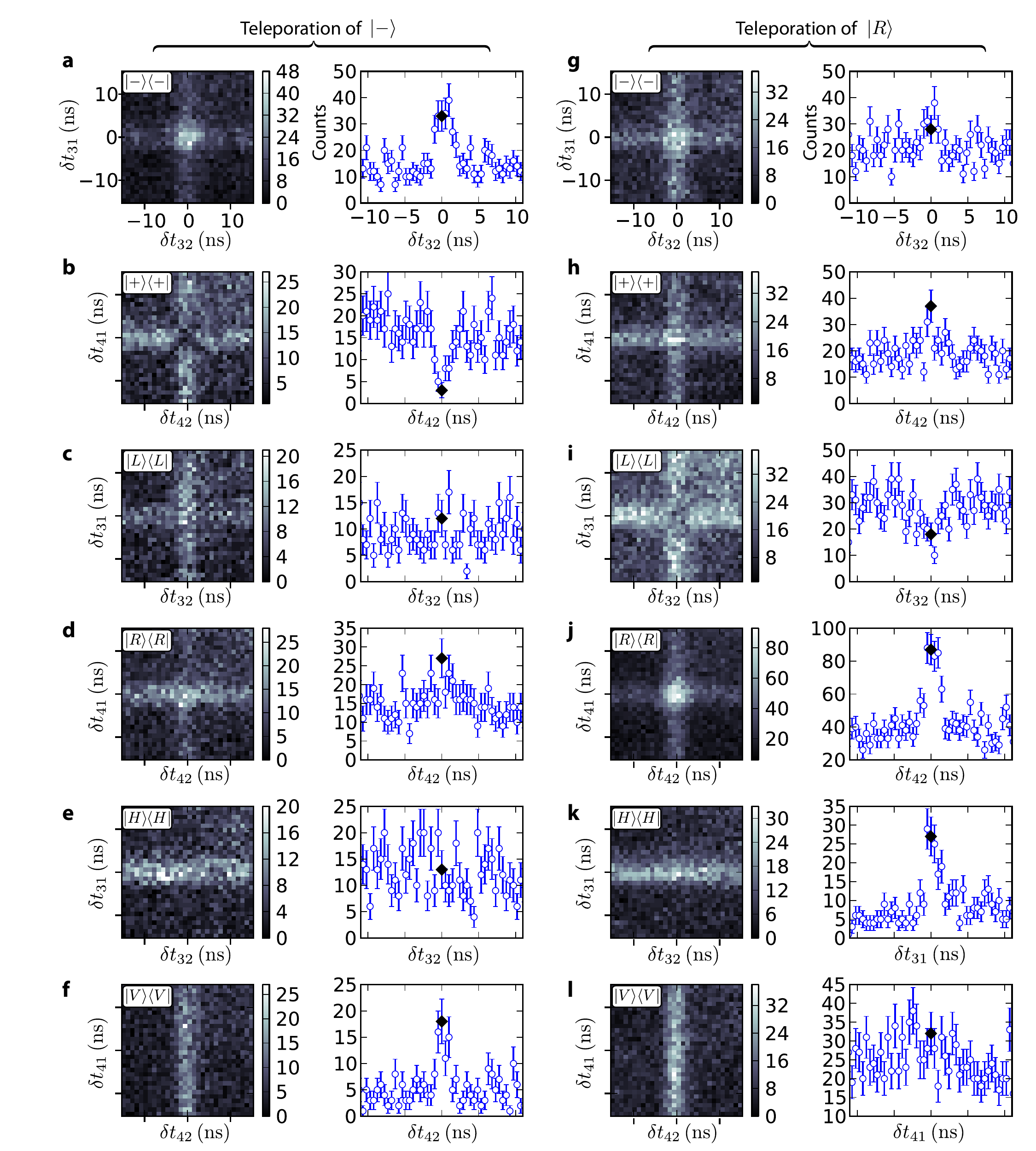}
\caption{Histograms of the number of threefold coincidences with the stored signal photon. They are shown as a function of delays $\delta t_{ij}$ between detector $D_i$ of the analyzer ($i=3,4$) and detector $D_j$ ($j=1,2$), where $D_j$ is one of the two SNSPDs used for the Bell state measurement. \textbf{a} and \textbf{b} show histograms corresponding to the teleportation of $\ket{-}$, when the analyzer was set to measure in the $\{\ket{+},\ket{-}\}$ basis. For \textbf{a} (or \textbf{b}), the left histogram is a shows the number of threefold coincidences with $D_1$, $D_2$ and $D_3$ ($D_1$, $D_2$ and $D_4$) or  as a function of $\delta t_{31}$  and $\delta t_{32}$ (or $\delta t_{41}$  and $\delta t_{42}$). Each pixel corresponds corresponds to a $(486~\text{ps})^2$ window. The one-dimensional histogram on the right side of \textbf{a} (or \textbf{b}) is a horizontal slice, centred on $\delta t_{31}=0$ (or $\delta t_{41}=0$),  of the associated two-dimensional histogram. \textbf{c} and \textbf{d} (or \textbf{e} and \textbf{f}) are the histograms corresponding to the teleportation of $\ket{-}$, when the analyzer was set to measure in the $\{\ket{R},\ket{L}\}$ basis ($\{\ket{H},\ket{V}\}$ basis). \textbf{g} through \textbf{l} are the histograms corresponding to the teleportation of $\ket{R}$. The black diamonds shown on the one-dimensional histograms are the points that have been used for the quantum state tomography.} \label{suppfig:echo}
\end{center}
\end{figure*}

\begin{figure*}
\begin{center}
\includegraphics[width=16.5cm]{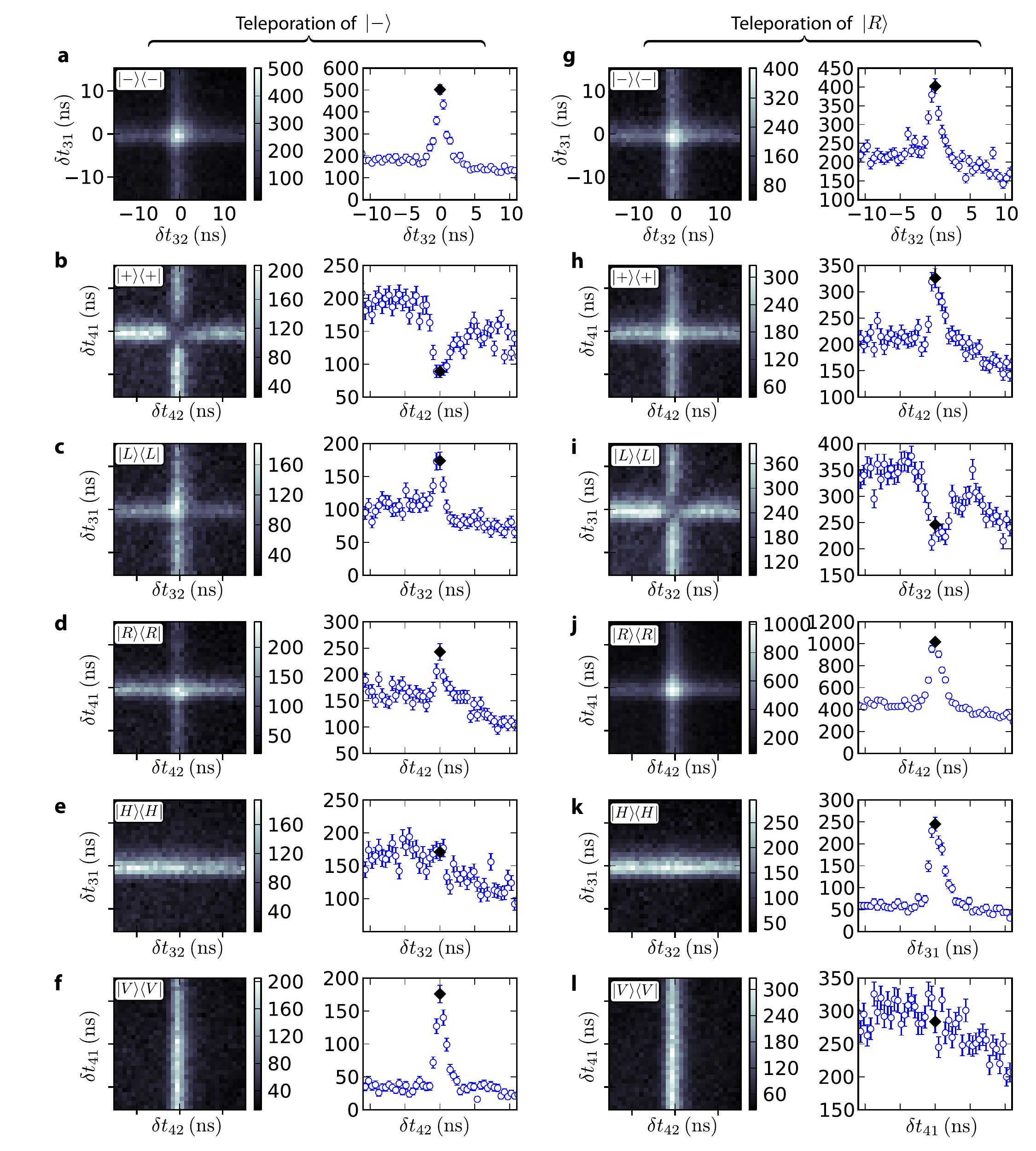}
\caption{Histograms of the number of threefold coincidences with the transmitted signal photon. They are shown as a function of delays $\delta t_{ij}$ between detector $D_i$ of the analyzer ($i=3,4$) and detector $D_j$ ($j=1,2$), where $D_j$ is one of the two SNSPDs used for the Bell state measurement. \textbf{a} and \textbf{b} show histograms corresponding to the teleportation of $\ket{-}$, when the analyzer was set to measure in the $\{\ket{+},\ket{-}\}$ basis. For \textbf{a} (or \textbf{b}), the left histogram is a shows the number of threefold coincidences with $D_1$, $D_2$ and $D_3$ ($D_1$, $D_2$ and $D_4$) or  as a function of $\delta t_{31}$  and $\delta t_{32}$ (or $\delta t_{41}$  and $\delta t_{42}$). Each pixel corresponds corresponds to a $(486~\text{ps})^2$ window. The one-dimensional histogram on the right side of \textbf{a} (or \textbf{b}) is a horizontal slice, centred on $\delta t_{31}=0$ (or $\delta t_{41}=0$),  of the associated two-dimensional histogram. \textbf{c} and \textbf{d} (or \textbf{e} and \textbf{f}) are the histograms corresponding to the teleportation of $\ket{-}$, when the analyzer was set to measure in the $\{\ket{R},\ket{L}\}$ basis ($\{\ket{H},\ket{V}\}$ basis). \textbf{g} through \textbf{l} are the histograms corresponding to the teleportation of $\ket{R}$. The black diamonds shown on the one-dimensional histograms are the points that have been used for the quantum state tomography.} \label{suppfig:transmission}
\end{center}
\end{figure*}

Let us consider the conceptual setup of Fig.~\ref{suppfig:simple-model} representing the teleportation of the $\ket{-}$ state. We will use it to explain the main features of the two 2D histograms corresponding to teleportation of the $\ket{-}$ state towards the signal, when the analyzer is set to measure in the $\{\ket{+},\ket{-}\}$ basis; see ~Fig.~\ref{suppfig:echo}-a and Fig.~\ref{suppfig:echo}-b (they are identical to Fig.~2-a and 2-b on the main text). For comparison, the histograms corresponding to the events where the detected signal photon was not absorbed by the memory (the transmitted photon) are shown on Fig.~\ref{suppfig:transmission}-a and \ref{suppfig:transmission}-b.
 Each histogram shows either the number of threefold coincidences at $D_1$, $D_2$ and $D_3$, or at $D_1$, $D_2$ and $D_4$. Each bin (i.e.~each pixel) corresponds to a window of fixed width and height, which is $(486~\text{ps})^2$ here. For a histogram with a detection at $D_j$ ($j=3$ or $4$), the $y$-axis corresponds to the delay $\delta t_{j1}$ between the detections at $D_j$ and $D_1$, and the $x$-axis to the delay $\delta t_{j2}$ between the detections at $D_j$ and $D_2$.

We first describe what we would expect in the vicinity of the central bin of the histogram, at $\delta t_{j1} = \delta t_{j2} = 0$, assuming ideal conditions (i.e~perfect optical alignment; negligible contribution from multi-photons in the WCS, multi-pairs and dark counts; negligible detection jitter and dark counts). This centre region corresponds to the threefold coincidences where the idler photon and a photon from the WCS were temporally overlapping at the 50/50 beam splitter (which heralds a successful Bell state measurement), and the detected signal photon is the entangled companion of the detected idler photon. The area of the region is of the order of $\tau_{i}^2$, where $\tau_i \approx 1.4$~ns coherence time of the idler photon. In this region, the probability to have a photon from the WCS just before the BS and to have an idler photon just before the BS, is given by $p\mu$ (we do not need to take into account the losses and detector efficiencies in the system because they all factor out in the final step of the calculation when we compare the probabilities for the different events). Given this, the probability that they split at the BS can be shown to be equal to $1/4$, which corresponds to the probability of a successful projection on the $\ket{\Psi^-} = 2^{-1/2}(\ket{HV}-\ket{VH})$ Bell state~\cite{Zukowski1993a}. Because the two photons are indistinguishable, they must have orthogonal polarizations behind the BS (otherwise they would bunch), but there are two possibilities happening with equal probabilities, namely $V$ in one mode and $H$ in the other, or the opposite. Hence, the presence of the orthogonally oriented polarizers after the BS  further reduces by a factor of~2 the probability to find one photon in each output arm after the polarizers. In practice, the polarizers were introduced to minimize the probability to detect two photons with the same polarization after the BS, which can happen if the photon are not perfectly indistinguishable. 
 
When a successful teleportation occurs, the polarization state of the signal photon is equal to the polarization state of the photon from the WCS, up to a constant unitary transformation that we include in the analyzer. We assume the analyzer is oriented such that a detection at $D_3$ corresponds to a projection on the input state. The total threefold coincidence probability is given by
\begin{equation}
P_{123}(\delta t_{31}=0,\delta t_{32}=0) = p \mu \cdot \frac{1}{4} \cdot \frac{1}{2} = \frac{1}{8} p \mu. \label{suppeq:goodprob}
\end{equation}
Since $D_4$ projects on $\ket{+}$, the probability to register a threefold coincidence at $D_1$, $D_2$ and $D_4$ should be zero:
\begin{equation}
P_{124}(\delta t_{41}=0,\delta t_{42}=0) = 0. \label{eq:probcenter}
\end{equation}

Let us now consider the case where the WCS photon is arriving later (by a time greater than $\tau_i$) compared to the two entangled photons. A possible realization of this would be a threefold with $\delta t_{31} = 0$ and $\delta t_{32} > \tau_i$. In this specific case, the detection at $D_2$ must stem from the WCS, because its detection time is not correlated to the detection of the signal photon, contrary to the idler photon (this is true only if $\mu \gg p$, which is shown in section~\ref{section:multi} to be a necessary condition to get a good signal-to-noise ratio). The probability for this threefold event is easily calculated to be
\begin{equation}
P_{123}(\delta t_{31}=0,\delta t_{32}>\tau_i) = \frac{1}{32}p\mu,
\end{equation} 
and we see that
\begin{eqnarray}
\frac{P_{123}(\delta t_{31}=0,\delta t_{32}=0)}{P_{123}(\delta t_{31}=0,\delta t_{32}>\tau_i)} = 4.
\end{eqnarray}

The same result applies to all the threefold events where one delay is zero, and the absolute value of the other is greater that $\tau_i$, i.e. for
\begin{eqnarray*}
& &P_{123}(\delta t_{31}=0,|\delta t_{32}|>\tau_i),\\ 
& &P_{123}(|\delta t_{31}|>\tau_i,|\delta t_{32}|=0),\\
& &P_{124}(|\delta t_{41}|>\tau_i,|\delta t_{42}|=0),\\
& &P_{124}(\delta t_{41}=0,|\delta t_{42}|>\tau_i).
\end{eqnarray*}

Finally, event with $|\delta t_{j1}|>\tau_i$ and $|\delta t_{j2}|>\tau_i$ correspond to threefold detections involving the creation of two entangled photon pairs created at different times and the detection of a WCS photon at a another time that differs from the previous two. These events happen with a probability of order $p^2\mu/32$ and are thus much less frequent than all the other ones consider above.

This simple model explains the 2D histograms of Fig.~\ref{suppfig:echo}-a-b with the stored photon, on which we see a cross-like structure centred on the origin, and either a peak or a dip at the centre. The one-dimensional histogram of Fig.~\ref{suppfig:echo}-a corresponds to a horizontal slice of the 2D histogram with $\delta t_{31} = 0$. We see a peak whose height rises above 30 counts, while the average number of counts away from the centre is approximately 12 per bin. The width of the peak is of the order of 2~ns, which is consistent with $\tau_i = 1.44$~ns. Similarly, we see a dip on the 1D histogram of Fig.~\ref{suppfig:echo}-b that nearly reaches 0 counts. The same structure appears for the teleportation of the $\ket{R}$ state with the analyzer set to the $\{\ket{R},\ket{L}\}$ basis (Fig.~\ref{suppfig:echo}-i-j). Finally, we also see the structure on the histograms of the transmitted signal photon, Fig.~\ref{suppfig:transmission}-a-b-i-j. Since all histograms obtained with the transmitted photon have more counts, they should be considered as indicators of what the results with the stored photons should be with better statistics. We note that the results for the teleportation of $\ket{+}$ are not shown here, but the above observations also apply for this state.

We now follow the same reasoning as above to explain the structure of the teleportation of $\ket{-}$, but when the analyzer is set to project in the $\{\ket{R},\ket{L}\}$ basis instead of $\{\ket{+},\ket{-}\}$ (we recall that this measurement is used in the quantum state tomography of the state retrieved from the memory). For the events that are away from the centre of the histogram, the probabilities are easily found to be the same as the previous case. For the events at the centre, the probability to detect the signal photon at $D_3$ must be half of the probability $P_{123}(\delta t_{31}=0,\delta t_{32}=0)$ calculated above (eq.~\ref{eq:probcenter}). This is because the retrieved state is in the $\ket{-}$ state and is analyzed in a maximally conjugated basis, which yields a 50\% probability of detecting it in a given detector. Hence, the 2D histograms corresponding to a detection at $D_3$ or $D_4$ should both show a peak at the centre, but with a height that is twice higher than the value found away from the centre. Fig.~\ref{suppfig:echo}-c-d show the relevant 2D histograms for the stored photon, on which the expected structure does not clearly appear, but the results are nevertheless conclusively different from the ones of Fig~\ref{suppfig:echo}-a-b. The expected structure is more apparent for the teleportation of the $\ket{R}$ state, when measured in the $\{\ket{+},\ket{-}$ basis (Fig.~\ref{suppfig:echo}-g-h). Finally, the expected structure clearly appears for the transmitted photon (Fig.~\ref{suppfig:transmission}-c-d-g-h).

The structure of the teleportation of $\ket{-}$, $\ket{R}$ and $\ket{+}$ when measured in the $\{\ket{H},\ket{V}\}$ basis is explained as follows. Let us assume the analyzer is set such that $D_3$ projects on $\ket{H}$. Due to the polarization entanglement between the signal and idler, a detection at $D_3$ remotely prepares the idler in the state $\ket{H}$. Therefore, the latter can only be detected at $D_1$ with a delay $\delta t_{31}=0$, which means that the coherent state can only be detected at $D_2$ to create a threefold coincidence. The probability for this process is $p\mu/16$, and it does not depend on the time at which the coherent state is detected. Because $D_1$ and $D_3$ are both projecting on $\ket{H}$, the probability to observe a coincidence between these two events with a delay $|\delta t_{31}|>\tau_i$ can only stem from the creation of more than one pair of entangled photons, and a threefold in this case would scale as $p^2 \mu$ and is much less probable. The structure of the 2D histogram should therefore consists in a peak centred on $\delta t_{31} = 0$ that is extending over all values of $\delta t_{32}$. Alternatively, the 2D histogram of $D_4$ should be a peak centred on $\delta t_{32}$ that is extending over $\delta t_{31}$. That is indeed what we observe with the stored photon (Fig.~\ref{suppfig:echo}-e-f-k-l), and with the transmitted photon (Fig.~\ref{suppfig:transmission}-e-f-k-l).

\subsection{Additional remarks}
We note that the measurement time required to generate each of the histograms on Fig.~\ref{suppfig:echo} and \ref{suppfig:transmission} typically varied from 8 to 15 hours. These variations partly explain why the number of counts in the crosses is not always the same. The other contribution is the different collection and detection efficiencies of $D_3$ and $D_4$. 

The time required to produce each histogram could have been reduced by a factor of 4 by using two more detectors placed on the unused ports of the polarizing beam splitters, which are after the BS (see Fig.~1 of the main text). This would allow the projection of the idler and the WCS photon on the $\ket{\Psi^+} = 2^{-1/2}(\ket{HV}+\ket{VH})$ Bell state with a probability of 1/4 (in the ideal case), and thus doubling the probability to herald a Bell state measurement~\cite{Zukowski1993a}.

\subsection{Multiphoton emission} \label{section:multi}
The setup described on Fig.~\ref{suppfig:simple-model} is also useful to estimate the contribution of the emission of more than one photon in the WCS, or the creation of more than one photon pair, to the noise. 

For the teleportation of $\ket{-}$ considered above, we can calculate the probability $P(2,0)$ of getting a threefold coincidence stemming from two photons in the WCS, while the idler photon is lost and the signal photon is detected (the calculation applies to the other states as well). This probability is given by the probability $\mu^2/2$ to have two WCS photons; the probability $p$ to create a pair; the probability $(1-\eta_i)$ to loose the idler photon on the path from the source to the BS; the probability $1/8$ for both WCS photons to split at the BS and to pass through the polarization filters; the probability $1/2\cdot \eta_s$ to detect the signal photon in a given detector of the analyzer with a transmission $\eta_s$. This amounts to 
\begin{equation}
P(2,0)=p\mu^2(1-\eta_i)\eta_s/32.
\end{equation}

We can also calculate the probability $P(0,2)$ of a threefold stemming from the detection of two idler photons, while the WCS photon is lost. For this, we need to evaluate the probability to create two pairs of photons and that the two idlers have orthogonal polarizations. The Hamiltonian $H$ of spontaneous parametric downconversion process with polarization entanglement is proportional to $H \sim \frac{1}{\sqrt{2}}(\hat{a}_H^{\dagger}\hat{b}_H^{\dagger} + \hat{a}_V^{\dagger}\hat{b}_V^{\dagger})$, where $\hat{a}^{\dagger}_H$ (or $\hat{b}^{\dagger}_H$) creates a photon in the idler mode (or the signal mode) with horizontal polarization, etc. The state we get when two pairs of photons are created is proportional to $\frac{1}{2}pH^2\ket{00}$, i.e.
\begin{equation}
\frac{p}{4}\left[(\opdag{a}_H\opdag{b}_H)^2 +  (\opdag{a}_V\opdag{b}_V)^2  + 2\opdag{a}_H\opdag{a}_V\opdag{b}_H\opdag{b}_V\right]\ket{00}.
\end{equation}
The probability to get two idler photons with orthogonal polarizations, and the same for the two signal photons ($\opdag{a}_H\opdag{a}_V\opdag{b}_H\opdag{b}_V$) is therefore $p^2/4$, and the probability that they split at the BS and get transmitted through the polarizer is $1/4$. When this is the case, the probability for one of the two signal photons to be detected in a given detector of the analyzer is $\eta_s/2$, which we multiply by two because we have two photons. Overall, the probability of this threefold is 	
\begin{equation}
P(0,2) \approx \left(1 - \mu -\frac{\mu^2}{2}\right)\cdot \frac{1}{32} \cdot p^2\eta_i^2 \cdot 2\eta_s \approx \frac{1}{16} \cdot p^2\eta_i^2\eta_s.
\end{equation}

Using similar arguments, we can calculate the probability of other processes contributing to the noise, and show that only the ones given above are significant. Let $P(1,1) = p\mu\eta_i/8$ be the probability to register a  threefold corresponding to an actual teleportation (see eq.~\ref{suppeq:goodprob}). In order to maximize the signal-to-noise ratio, we must satisfy $P(1,1)\gg P(2,0)$ and $P(1,1)\gg P(0,2)$, which translates to
\begin{eqnarray*}
\eta_i &\gg& \frac{\mu}{4+\mu} \approx \frac{\mu}{4}, \\
\mu &\gg& \frac{1}{2}\cdot p\eta_i.
\end{eqnarray*}
By combining the two inequalities we get
\begin{equation}
\eta_i \gg \frac{\mu}{4} \gg \frac{p\eta_i}{8}. \label{suppeq:condition}
\end{equation}
The measured values are $\eta_i \approx 0.13$, $\eta_s \approx 6.3\times 10^{-3}$, $p\approx 10^{-2}$ and $\mu \approx 0.011$, which satisfy the inequality. The value of $p$ given here is taken as $1/g_{si}^{(t)} \approx 1/100$.

\section{Quantum state tomography} \label{section:tomography}
We performed quantum state tomography to obtain complete information about the state retrieved from the memory. Quantum state tomography can be performed by measuring the photon in the three usual bases ($\{\ket{H},\ket{V}\}$, $\{\ket{+},\ket{-}\}$ and $\{\ket{R},\ket{L}\}$), from which the $x$,$y$ and $z$ components of the Bloch vector are extracted~\cite{Altepeter2006a}. Measurements with the different bases was performed in an alternating fashion, i.e.~we would measure in each basis for one hour, and then cycle through. Quantum state tomography requires a suitable normalization of the observed number of counts to compensate for the uneven detection efficiencies of $D_3$ and $D_4$. The method used is based on the following reasoning. 
Since the state of the signal photon sent to the polarization analyzer after the quantum memory is from a close-to-maximally entangled pair,  its polarization is in the completely mixed state (when we trace out the idler photon). The probability to observe it in $D_3$ or $D_4$ is therefore be same, provided their collection and detection efficiencies are identical. Any observed deviation can therefore be used to normalized the number counts observed in a given period of time. To measure this deviation, we used the 2D histograms discussed in section~\ref{section:results}, but with delays extending from -100 to 100~ns instead of -15 to 15~ns. Any threefold coincidence for which the delays are not equal to each other (i.e.~away from the diagonal) heralds a signal photon in a completely mixed state. Hence, counting the total number of such events for the histogram with a detection at $D_3$, and comparing it to the number extracted from the equivalent histogram for $D_4$, directly gives the information about the detection efficiency mismatch. It can then be used to normalize the number of counts. This method is reliable since the wide area covered in the histogram yields a statistically significant number of counts, from which a good estimate of the mismatch is obtained. This method works only if the source is well balanced (that is the probability to find the pair in $\ket{HH}$ is essentially the same as finding it in $\ket{VV}$). This condition was satisfied in our experiment, and was automatically checked every hour.  

The above method does not work for the teleportation of $\ket{H}$, because here the state of the signal photon is, ideally, in a pure polarization state and the aforementioned argument does not work. The efficiency mismatch was instead extracted from the maxima of the fitted visibility curves that are measured every hour. 

The delays $\delta t_{ij}$ ($i=3,4$ and $j=1,2$) of Fig.~\ref{suppfig:echo} and \ref{suppfig:transmission} have all been adjusted to position the events corresponding to an actual teleportation at the centre of the histograms. To determine the offsets that needs to be applied, we used the data accumulated for the visibility curves and from which we produced histograms of the twofold coincidences between the four detection combinations (1-3, 1-4, 2-3 and 2-4). For each histogram, the stored photon peak was fit with a gaussian, and the position of the maxima was used to adjust the offsets of the histograms. 

\subsection{Fidelity and purity}\label{section:fidelity}
The fidelity $F$ of a mixed state $\rho$ with respect to a pure target state $\ket{\psi}$ is defined as $F = \bra{\psi}\rho\ket{\psi}$. It corresponds to the probability obtaining the result $\ket{\psi}$ when subjecting $\rho$ to a projective measurement in the orthonormal basis $\{\ket{\psi}, \ket{\psi^{\bot}}\}$, where $\braket{\psi}{\psi^{\bot}}=0$. In practice, the effect of loss in the channel is post-selected out by keeping only the events where the signal photon is detected. For our experiment, the input states were always contained in one of the measurement bases used for the quantum state tomography of the retrieved state. Hence, the measurement in that basis can readily be used to estimate the fidelity. Let $N_3$ (or $N_4$) be the number of threefold coincidences observed at $D_3$ (or $D_4$), properly normalized to compensate for its efficiency mismatch (see section~\ref{section:tomography}). If $D_3$ is the detector projecting on the target state, then the fidelity is directly given by $N_3/(N_3+N_4)$. It can also be written as $F=(1+V)/2$, where $V = (N_3-N_4)/(N_3+N_4)$ is called the visibility of the state. 

 \begin{table*}[!t]
 \caption{Measured values of the fidelity and the purity for different input states, with either the stored or the transmitted signal photon. The purity-limited fidelity $F_{\text{max}}$ is also given for the stored photon, for comparison with the measured fidelity.} \label{supptable:results}
\begin{ruledtabular}
 \begin{tabular}{llcrlcr}
      & \multicolumn{3}{c}{\textbf{Stored photon}} & \multicolumn{3}{c}{\textbf{Transmitted photon}} \\
    State & Purity (\%) & Fidelity (\%) & $F_{\text{max}}$ & Purity (\%) & Fidelity (\%) & $F_{\text{max}}$ \\ 
    \hline
    $\ket{H}$ 	& $93\pm 5$ 	& $94\pm 3$ 	& $96\pm3$ 	& $94\pm 1$ 	& $95\pm 1$ 	& $97\pm1$	\\ 
    $\ket{-}$  	& $94\pm 6$ 	& $92\pm 4$ 	& $97\pm 3$ 	& $77\pm 2$ 	& $86\pm 1$ 	& $87\pm 1$	\\
    $\ket{R}$ 	& $73 \pm 5$	& $82\pm 4$ 	& $84\pm 4$ 	& $69 \pm 1$ 	& $79\pm 1$ 	& $81\pm1$	\\
    $\ket{+}$ 	& $83 \pm 9$	& $82\pm 4$ 	& $91 \pm 6$ 	& $68 \pm 2$ 	& $80\pm 1$ 	& $80\pm 2$	\\
    $\ket{+}$ (12.4~km) &  --- & $81\pm 4$ & --- &  --- &$80 \pm 1$ & --- \\
  \end{tabular}
\end{ruledtabular}
\end{table*}

The previous measurement is combined with the measurements in the other two bases to construct the Bloch vector $\vect{r}=r_x \uvect{x} + r_y\uvect{y}+r_z\uvect{z}$ of the retrieved state, which is used to parametrize the state $\rho$ as $\rho = (1+\vect{r}\cdot \bsigma)/2$, where $\bsigma = \sigma_x \uvect{x} +\sigma_y \uvect{y} + \sigma_z \uvect{z}$ is the vector of Pauli matrices. It can also be used to estimate the purity $P$ of the state, defined as $P=\text{Tr}(\rho^2)$. It is related to the length of the Bloch vector through
\begin{equation}
P = \frac{1}{2}(1+|\vect{r}|^2).
\end{equation}
The purity should be equal to~1 if the experimental noise is negligible, and $1/2$ if we measure unbiased noise only. Hence, the purity is an indicator of the signal-to-noise ratio of the teleportation itself. 

The fidelity decreases with the purity and with any unwanted rotation around the Bloch sphere, that could be due to, e.g.~optical misalignment. We could therefore get some indication about whether the less-than-unity fidelities we observed are mostly due to the purity reduction (assuming the input state is pure) or to a rotation. Specifically, let us assume there is no such rotation, and that the effect of the teleportation is to recreate the target state $\ket{\psi}$ with a probability $V$, mixed with white noise with a probability $1-V$: 
\begin{equation}
\rho = V\projector{\psi}{\psi} + (1-V)\frac{\mathbb{I}}{2}
\end{equation}
where $\mathbb{I}$ is the $2\times2$ identity matrix. In this model, the probability $V$ corresponds to the visibility of the state defined above. Using $F=(1+V)/2$, we have
\begin{equation}
\rho = F\projector{\psi}{\psi} + (1-F)\projector{\psi^{\bot}}{\psi^{\bot}}.
\end{equation}
In this case, the purity can be written as a function of the fidelity:
\begin{equation}
P(F) = \text{Tr}(\rho^2) = 2F^2 - 2F +1. \label{suppeq:purity}
\end{equation}
We can also write the fidelity as a function of the measured purity,
\begin{equation}
F_{\text{max}} = \frac{1}{2}(1+\sqrt{2P-1}).
\end{equation}
Inserting the measured value of the purity in the previous equation effectively yields an upper bound to our measured value of the fidelity. If the measured value of the fidelity is close to this upper bound, than we can say that it is mostly noise-limited.

Our experimental results for the teleportation to the stored photon are presented in Table~\ref{supptable:results}. The uncertainties are evaluated using Monte Carlo simulations assuming that the number of counts measured follows a Poisson distribution. We notice that the purity varies from state to state, which is an indication that the experimental parameters that affected the signal-to-noise ratio (see section~\ref{section:sourcecharacterization}) also varied during the measurements. We also notice that the measured fidelity is very close to the noise-limited upper bound $F_{\text{max}}$, except for $\ket{+}$. 

The results of the teleportation with the transmitted photon are shown on Table.~\ref{supptable:results}. The fidelities are all close to the results obtained with the stored photon. Close inspection however reveals that all the fidelities are slightly lower than their stored photon counterparts, and the difference is more important for the purities. The most likely explanation for this is related to the fact that storage acts as a temporal filter which selects only the stored light~\cite{McAuslan2012a} and effectively removes other spurious sources light. 

We can compare these values to what is expected from the model described in section~\ref{section:multi} (assuming there is no additional rotation of the Bloch vector). Specifically, the fidelity can be estimated from the expressions $P(1,1)$, $P(2,0)$ and $P(0,2)$ given above:
\begin{equation}
F = \frac{P(1,1)+P(2,0)+P(0,2)}{P(1,1)+2[P(2,0)+P(0,2)]}, \label{suppeq:fid}
\end{equation}
from which the purity can also be calculated using eq.~\ref{suppeq:purity}. With the experimental parameters $\eta_i \approx 0.13$, $\eta_s \approx 6.3\times 10^{-3}$, $p\approx 10^{-2}$ and $\mu \approx 0.011$, we find $F\approx 0.93$ and $P\approx 0.88$, which is close to what we measured. This gives an indication that multi-pairs and multi-photons are the main factors affecting the fidelities and purities that we measured.  

We note here that we assumed $p \approx 1/g_{si}^{(t)} = 1/100$, which would be correct if the transmission spectra of the idler and the signal filters had the same widths, but this was not the case. Hence, using that relation actually slightly overestimates the value of $p$. 

We also note that the measured values of $p \sim 1/g_{si}^{(t)}$ varied during our measurements, by approximately a factor of two at most (see section~\ref{section:gsi}). This contributed to the fluctuations of the fidelity and purity.

\subsection{Fidelity of the teleportation with the 12.4~km fibre spools}
We also performed a teleportation of the $\ket{+}$ state in a configuration where the WCS and the idler photon each travelled through 12.4~km of standard single mode optical fibres before the BSM; see Fig.~1 of the main text. The measured fidelity obtained from the measurement in the $\{\ket{+},\ket{-}\}$ basis is $81\pm 4\%$. It is of the same order as the one measured without the fibres, which is consistent with the fact that the loss introduced by the fibre spools is the same for the idler mode and the WCS. Specifically, the transmission $\eta$ of the fibres changes $\mu$ and $\eta_i$ to $\eta \mu$ and $\eta\eta_i$ in the inequality of Eq.~\ref{suppeq:condition}, which satisfies it just as well as the one without $\eta$ and leaves the expression of the fidelity of Eq.~\ref{suppeq:fid} unchanged (when neglecting other sources of noise such as dark counts).

\newpage

\newpage

\end{document}